\documentclass[12pt,preprint]{aastex}
\usepackage{color}
\usepackage{ulem}
\usepackage{epsfig}

\shorttitle{KHz QPOs from Terzan 4}
\shortauthors{Didier Barret}

\def\4u{4U1608-522}
\def\exo{EXO1745-248}
\def\igr{IGR J17480-2446}

\def\gsim{\mathrel{\hbox{\rlap{\hbox{\lower4pt\hbox{$\sim$}}}\hbox{$>$}}}}
\def\lsim{\mathrel{\hbox{\rlap{\hbox{\lower4pt\hbox{$\sim$}}}\hbox{$<$}}}}
\def\minchi2{min($\chi^2$)}

\begin{document}

\title{kHz QPOs from the 2000 and 2010 X-ray transients located in the globular cluster Terzan 5: \exo\ and \igr}

%
%
\author{Didier Barret\altaffilmark{1}}
\affil{Institut de Recherche en Astrophysique et Plan\'etologie \& Universit\'e de Toulouse (UPS), 9 avenue du Colonel Roche, BP 44346, 31028, Toulouse Cedex 4, France}

\email{didier.barret@irap.omp.eu}




\begin{abstract}
EXO1745-248 is a transient neutron star low-mass X-ray binary located in the globular cluster Terzan 5. It was in outburst in 2000 and displayed during one {\it Rossi X-ray Timing Explorer} observation  a highly coherent quasi-periodic oscillation (QPO) at frequencies between 670 and 715 Hz. Applying a maximum likelihood method to fit the X-ray power density spectrum, we show that the QPO can be detected on segments as short as $T=48$ seconds. We find that its width is consistent with being constant, while previous analysis based on longer segment duration (200 s) found it variable. If the QPO frequency variations in \exo\ follows a random walk (i.e. the contribution of the drift to the measured width increases like $\sqrt{T}$), we derive an intrinsic width of $\sim 2.3$ Hz. This corresponds to an intrinsic quality factor of $Q\sim 297\pm50$ at 691 Hz. We also show that $Q$ is consistent with being constant between 2.5 and 25 keV. 

\igr\ is another X-ray transient located in Terzan 5. It is a very interesting object showing accretion powered pulsations and burst oscillations at 11 Hz. We report on the properties of its kHz QPOs detected between October 18th and October 23rd, soon after the source had moved from the so-called Atoll to the Z state. Its QPOs are typical of persistent Z sources; in the sense that they have low $Q$ factors ($\sim 30$) and low RMS amplitudes ($\sim 5$\%). The highest frequency (at $870$ Hz), if orbital, sets a lower limit on the inner disk radius of $\sim 18.5$ km, and an upper limit to the dipole moment  of the magnetic field $\mu \le 5\times 10^{26}~{\rm G~cm}^3$.
\end{abstract}


\keywords{accretion, accretion disks, stars: neutron, X-rays: binaries, X-rays: stars}
%
%

\section{Introduction}
Terzan 5 is a galactic globular cluster, which contains several X-ray sources \citep{heinke03}, including \exo: an X-ray transient that was in outburst in 2000 and 2002 \citep{mar00,wij02}. A strong kHz QPO from \exo\ was discovered by \cite{mb11} from the RXTE data recorded during the 2000 outburst.  \cite{mb11} using Power Density Spectra (PDS) with an integration time of 10 s reported that when fitting the QPO on timescales down to 200 s, the quality factor of the QPO was variable and reached a peak value of $264\pm38.5$, despite the fact the no correction for the frequency drift was applied, within the 200 second intervals. \cite{mb11} thus argued that the shift-and-add procedure \citep{mendez98,barret05}, which in its general form involves  the average of a large number of PDS, could miss high $Q$ QPOs, appearing in segments of short durations. In their analysis \cite{mb11} suggested also that the quality factor of the QPO could increase with energy. In this paper, we revisit the above observation, by applying a Maximum Likelihood Estimation (MLE) of the QPO parameters, following the method described by \cite{bv11}. The MLE has been shown more sensitive and much less biased than the more traditionally used \minchi2, especially when the number of PDS fitted is small, as is the case when attempting to track the QPO frequency on short timescales.  In section \ref{data_analysis}, we present the observation and our analysis, which we complement with simulations to show the level of scattering expected in the fitted $Q$ values, even for a signal of a constant $Q$ factor.

\igr\ is another neutron star X-ray transient discovered in 2010 in the globular cluster Terzan 5. It was originally detected by the INTEGRAL/IBIS imager \citep{bordas10,ferrigno10}. Investigation of archival Chandra data revealed that \igr\ was distinct from \exo\ \citep{heinke}.  \igr\ displays very interesting properties: it is an 11 Hz eclipsing bursting X-ray pulsar \citep{chenevez10,strohmayer10,linares10,papitto11,linares11}, showing burst oscillations locked with the accretion powered pulsations \citep{altamirano10a,cavecchi11}. The spectral evolution along its outburst indicated that it made a transition from the so-called Atoll state to the Z state, around October 16-17th \citep{altamirano10b,chakraborty11}. \cite{altamirano10b} reported a marginal detection of a kHz QPO at about 815 Hz ($3.8\sigma$, single trial significance, 10-50 keV, October 18th), but so far no systematic search for high frequency QPOs from \igr\ has been presented. This is the scope of section 3, where we report on highly significant high frequency QPOs from \igr.  We discuss the results on both sources in section 4.

\section{Revisiting the QPO from \exo}
\label{data_analysis}
The 2000 outburst of \exo\ was monitored by RXTE between July and November, 2000. We have extracted Leahy normalized Power Density Spectra (PDS), using all events recorded between 3 and 25 keV and an integration time of 8 s (Nyquist frequency at 2048 Hz). Two proposals P50054 and P50138 are considered, amounting to 126 ks of observations. As reported by \cite{mb11}, in one segment only, a significant kHz QPO was detected (ObsID: 50054-06-11-00, starting on 2000-09-30 at 15h 28 m 53 s, and lasting for 3230 seconds). This observation was taken in the tail of the outburst decay, about 40 days after the peak. The dynamical PDS is shown in Figure \ref{dynpds}. The track followed by the QPO can be easily seen, and its frequency varies from $\sim 713$ Hz and $\sim 672$ Hz, and displays relatively large jumps by up to 20-30 Hz on timescales of hundreds of seconds. 
\subsection{Measuring the QPO width}
In order to reduce the contribution of the frequency drift to the measured QPO width, we minimize the number of 8 s PDS to be averaged ($M$), and yet to have a significant detection of the QPO. The PDS is modeled as the sum of a constant $a$ (to account for the Poisson noise, i.e. equal to $2$ for Leahy normalized PDS, \cite{leahy83}), plus a Lorentzian accounting for the QPO, with three parameters: the normalization $R$ (equal to the integrated power of the Lorentzian from 0 to $\infty$), the width $w$ (FWHM) and the centroid frequency $\nu_0$. {The RMS amplitude is a derived quantity, computed as RMS=$\sqrt{R \times (S+B)/S^2}$ expressed in percentage, in which $S$ is the net source count rate and $B$ is the background count rate, $B$ is often negligible compared to $S$}. We define the minimum $M$ as to have more than 80\% of the intervals in which $R/\Delta R \ge 3$, where $\Delta R$ is the $1\sigma$ error on $R$ \citep[see e.g.,][]{boutelier10}. For this observation, we have found a minimum $M$ equal to 6, although $M=5$ yields a fraction of significant detections of 75\%. The QPO parameters can thus be recovered down to a 48 second timescale. The MLEs of the QPO parameters are shown in Figure 2 for $M=6$. The mean of the MLEs of $w$ ($<w^{MLE}>=3.8\pm0.2$ Hz) measured for $T=48$ seconds (the error is given as the standard error on the mean).  The mean value of $R/\Delta R$ is about 4, indicating that the biases on $R$ and $w$ are negligible (the leakage bias at $T=8$ s is only 0.04 Hz, 50 times less than the error bar on $<w^{MLE}>$, \cite{bv11}). For a mean QPO frequency of 690 Hz, this width corresponds to a mean quality factor of $Q\sim182$, or $Q=177, 189 \pm 8$ at the minimum and maximum frequencies detected. Over 48 s, $Q$ ranges from $82\pm 29$ to $754\pm 390$. Fitting of the $\log w^{MLE}$ values \cite[closer to a normal distribution, see][]{bv11} with a constant yields a $\chi^2$ of 53.4 for 57 degrees of freedom. Hence, within the statistical quality of the data, $w$ is consistent with being constant over the observation, despite the frequency variations, not corrected for within the 48 s of the cumulative PDS integration time. We have reprocessed the data with $M=25$, corresponding to a cumulative integration time of 200 s for the PDS, as in \cite{mb11}. $<w^{MLE}>$ becomes $5.9\pm0.5$ Hz, and is not consistent with being constant (due to the fact that not all intervals experience the same frequency drift and the contribution of the drift to the measured width becomes significant). For a mean QPO frequency of 690 Hz, this corresponds to a mean quality factor of $Q\sim 118$, or $Q=115, 121 \pm 10$ at the minimum and maximum frequencies detected. Over 200 seconds, $Q$ ranges from $64\pm 13$ to $281\pm 65$; the latter value being consistent with the maximum value reported by \cite{mb11}. 

The effect of increasing $M$ or the segment duration is obviously to increase $<w^{MLE}>$ and to increase the scatter due to the variable frequency drift. This is shown in Figure \ref{rw} where we show  how $<w^{MLE}>$ increases with $T=M\times8$ s ($M$ ranges from 5 to 12). As discussed in \cite{bv11}, if the frequency drift follows a random walk, we expect the broadening of the QPO due to the variable frequency to increase with $\sqrt{T}$.  The power spectrum of the frequency variations, using the estimates of the QPO frequency on a timescale of 48 s can be adequately fitted by a simple power law model of index $2.6 \pm 0.6$, and is therefore consistent, within error bars, with a random walk (the index should be 2). Assuming that the measured width for a given $T$ ($w_T$) includes a contribution from the intrinsic QPO width $w_{qpo}$ and a contribution from the drift ($w_{drift}$), such as $w_T^2=w_{qpo}^2+w_{drift}^2$, with $w_{drift} \propto \sqrt{T}$, it is possible to recover the intrinsic width of the QPO. As can be seen from Figure \ref{rw}, the random walk model can fit the data. We found $w_{qpo}=2.3\pm 0.4$ Hz or $Q_{qpo}=297\pm50$ at 691 Hz.

\subsection{Simulations}
As shown above, the measured width and its scatter both increase with $M$ or equivalently the interval duration, but on top of this, it should also be realized that fluctuations around the mean QPO profile will also scatter $Q$ around its mean value. This is best illustrated with a simulation of a set of synthetic PDS, containing a QPO with fixed frequency, amplitude and width. Following the method described in \cite{bv11}, we have simulated 1050 PDS of 8 s duration from segments  256 s long, containing a QPO with the parameters derived from \exo\ ($\nu_0=690$ Hz, RMS$_0=9.2$\%, $w_0=2.3$ Hz, equivalent to $Q_0=300$). Fitting the average of the 1050 PDS gave $w=2.3\pm0.1$ Hz, an RMS amplitude of $9.2\pm0.1$ \%, and quality factor $Q=295 \pm 7$, consistent with the input parameters. The MLEs of the QPO parameters derived for $M=25$ (equivalent to $T=200$ s) are shown in Figure \ref{mle_sim}. As can be seen, despite a constant QPO width and frequency (hence fixed $Q$), $Q$ shows some significant scattering, with minimum and maximum values of  $ 217\pm 31$ and $470 \pm 72$. Adding a variable frequency for the QPO would shift down all the $Q$ estimates by a similar amount on average. The scattering will remain, and for this reason, taking the maximum $Q$ value found as a measure of the underlying $Q$ factor of the QPO is not correct. At any given timescales, a better estimate of $Q$ is obtained as the average of many measurements \cite[as the shift-and-add technique does,][see also Figure 4]{barret06}.


\subsection{Q dependence with energy}
\cite{mb11}  suggested that the quality factor of the QPO could increase with energy, { although within error bars $Q$ was consistent with being constant}. No correction for the frequency drift was applied in their analysis, explaining why the $Q$ values of their Figure 5 are low (around 30). { First we failed to reproduce the smooth trend found in \cite{mb11}, extracting PDS in the same energy bands and using \minchi2\ fitting (or even ML fitting). Second, it should be stressed that to get meaningful $Q$ values, a correction for the frequency drift must be applied.} Using the time history of the QPO frequency as measured on 48 s, we have shifted-and-added all the PDS in four adjacent energy bands. The MLEs of the QPO parameters are shown in Figure \ref{qene}. As can be seen, within the statistical quality of the data, the quality factor of the QPO is consistent with being constant between 2.5 and 25 keV. The RMS amplitude of the QPO increases as it does in similar systems \citep[for 4U1608-525,][]{berger96}. 
\subsection{Q dependence with frequency}
It is known that the quality factor of the QPO varies with frequency, increasing smoothly with frequency up to a maximum value beyond which a sharp drop is observed \citep{barret06}. This effect was observed after averaging large data sets (to reduce the uncertainty in $Q$), when the QPO spans a large frequency interval (300-400 Hz wide). We have searched for such a variation in the data by grouping 100 8 s PDS of adjacent frequencies, shifting and adding them to the mean frequency. The MLEs of the QPO parameters as derived from the shifted-and-added PDS are presented in Figure \ref{sa100}. Clearly the statistical quality of the data does not allow to see variations of $Q$ (and the RMS amplitude) with frequency, over the 40 Hz interval covered by the observation.  
\section{kHz QPOs from IGR J17480-2446 during its 2010 outburst}
\igr\ was monitored during its 2010 outburst from  between October 13th and November 19th. We have processed all the archival data corresponding to proposal P95437 (total observing time of about 284 kiloseconds for 46 ObsIDs), and produced 8 s PDS, using all events of energies between 5 and 25 keV (this energy range maximizes the signal-to-noise ratio of the QPOs detected). No screening for X-ray bursts was performed. We split the ObsIDs in time continuous intervals; e.g. switching the PCA from 1 PCU to 2 PCUs defines an interval boundary. We have a total of 102 intervals.  A blind search for high frequency QPOs at frequencies above 500 Hz has been performed on the interval averaged PDS, using a scanning algorithm \citep{boirin00}, and a single trial significance threshold of the excess power of $5\sigma$, below which spurious features are often found \citep[see e.g.,][]{boutelier10}.

High frequency QPOs with significance larger than $6\sigma$ (single trial) were  detected within four ObsIDs: 95437-01-06-000 ($7.6\sigma$), 95437-01-07-00 ($7.1\sigma$), 95437-01-09-00 ($12.1\sigma$) and 95437-01-10-01 ($8.2\sigma$). { The ObsID 95437-01-06-00, in which \cite{altamirano10b} reported a marginal QPO detection is split in two intervals. No excess is found above $5\sigma$ (even $4\sigma$) in either intervals or in their combination. The same conclusion applies to the PDS extracted using events of energy between 10 and 50 keV, as \cite{altamirano10b} did. { The strongest excess reaches $3.5\sigma$ (single trial) in the 5-25 keV band (at $\sim 825$ Hz), combining the two intervals. Accounting for the number of trials of the scanning routine ($\sim 10^3$), this excess is clearly not significant. We therefore conclude that no significant QPOs were detected in ObsID 95437-01-06-00.} 

The high frequency QPOs we have detected have been fitted with the Maximum Likelihood on timescales of 1024 s for the three intervals and 512 s for the second interval of the ObsID 95437-01-09-00. The maximum likelihood estimates of the QPO parameters are reported in Table 1. As can be seen, all the detections occurred soon after the source had transited to the Z state. Not surprisingly, its QPOs are relatively broad (few tens of Hz) and of low RMS amplitude (3-6 \%), which are characteristic properties of QPOs of Z sources  \citep{klis00}. We have shifted-and-added all the PDS containing a high frequency QPO: at the mean frequency of 839 Hz, the quality factor of the QPO is $Q=$32.9$\pm$3.7 and the RMS amplitude is 4.7$\pm$0.2 \%. No other features were present in the averaged PDS between 500 and 1500 Hz, with an upper limit on the RMS amplitude of about 2\% for a signal of 20-50 Hz width. 

\section{Discussion}
We now discuss the results obtained above, starting first with \exo. 
\subsection{\exo}
We have re-analyzed the data from EXO1745-248, in which \cite{mb11} discovered a single high frequency QPO, most likely to be a lower kHz QPO, due to its high quality factor and amplitude. We have applied the recently developed Maximum Likelihood method to fit the QPO on the shortest timescales permitted by the data statistics; 48 s in this case. Contrary to \cite{mb11}, we have found that the QPO width is consistent with being constant within the observation, corresponding to a mean quality factor of $\sim 182$ at 690 Hz (on 48 s). We have failed to detect variations of $Q$ with frequency within the 40 Hz frequency interval sampled by the QPO, but this is again likely due to the limited signal to noise ratio of the data. It is therefore impossible to tell whether $Q$ from \exo\ could still reach higher values. Within the statistical quality of the data, we have shown that Q, corrected for the frequency drift on timescale of 48 s, does not depend on energy. Applying a simple random walk model, we have inferred that the intrinsic quality factor of the QPO could be as large as $297\pm50$ at 691 Hz (Fig. 3). Despite the relatively large error bar, this puts \exo\ amongst the QPO sources with the most coherent QPOs, and therefore provides even stronger constraints on QPO models, due to the long inferred lifetime for the underlying oscillator \cite[see discussion in][]{barret05}. We have also shown with simulations that for the QPO parameters recovered for \exo, the $Q$ values measured on 200 s intervals display some significant scattering. Therefore techniques based on averaging many individual measurements (e.g. as with the shift-and-add technique) are appropriate for deriving meaningful quality factors.
\section{\igr}
We have reported high frequency QPOs from \igr: another X-ray transient located in Terzan 5, which was in outburst in late 2010. First, we failed to confirm the marginal QPO detection previously reported by \cite{altamirano10b}, but we detected highly significant QPOs  between October 18th and 23rd, at frequencies between 800 and 870 Hz. These QPOs were detected soon after the source had moved from the Atoll to the Z state (around October 16-17th). The QPOs are typical of Z  sources, in the sense that they are broad and have a relatively low amplitude. They are similar to those detected in the Z phase of the other system, which shows both Atoll and Z states, namely XTE J1701-462 \citep{homan07,sanna10}. Unlike in the case of  XTE J1701-462, no narrower high frequency QPOs were detected in the Atoll phase of \igr, while the sensitivity of the observations was sufficient (e.g. limiting RMS around 3\% for a QPO width of 5 Hz on October 14th), had the QPO been as strong as the one seen from XTE J1701-462 \citep[RMS amplitude between 8 and 10\%, see][]{sanna10,bbm11}. It is hard to tell whether the QPOs detected are lower or upper QPOs, because only a single one was detected in all intervals, and in the combined intervals. The analogy with XTE J1701-462 does not help, as \cite{sanna10} have shown that in its Z phase, the upper and lower QPOs had similar $Q$ and RMS amplitudes. In general though, in classical Z sources, e.g. Sco X-1, the lower and upper QPOs have comparable $Q$ factors, but the upper QPOs have larger amplitudes and should therefore be easier to detect \citep{boutelier10}. Note however, that in \igr\ the spin frequency of the neutron star ($\nu_s$) is 11 Hz, and that in many systems, the frequency separation of the QPOs is either found close to $\nu_s$ or $\nu_s/2.$ \citep[see however,][]{mendez07}: detecting two QPOs with such a separation would require very narrow (and comparatively strong) QPOs to start with; this could explain why only one single QPO is seen from \igr\ in its Z phase. 

The highest frequency at 870 Hz was detected when the source X-ray luminosity was $\sim 4 \times 10^{37}$ ergs/s \citep{chakraborty11}, at the distance of 5.5 kpc \citep{ortolani07}. Assuming that this frequency is an orbital frequency at the inner disk radius ($R_{in}$); this sets a lower limit on $R_{in}$ as $\sim 18.5 m_{1.4}^{1/3}$ km (where $m$ is the NS mass, given in units of 1.4 $M_\odot$). According to \cite{burderi01}, the magnetospheric radius $R_m$ is given as $R_m \sim 160 ~ m_{1.4}^{1/7} R_6^{-2/7} L_{37}^{-2/7} \mu_{28}^{4/7}$ km, where $R_6$ is the NS
radius in units of 10 km, $L_{37}$ the accretion luminosity in units of $10^{37}$ erg s$^{-1}$, and $\mu_{28}$ the magnetic dipole moment of the NS in units of $10^{28}$ G cm$^3$. Assuming that the disk is truncated at the magnetospheric radius, an upper limit $\mu \le 5\times 10^{26}~{\rm G~cm}^3$ can be derived on the magnetic dipole moment of the NS (for canonical parameters for the neutron star). Under the simplistic assumption that  $\mu=B_s R^3$, the surface magnetic field would be less than $5\times 10^{8}~{\rm G}$, a value at the lower end of the range derived previously from the detection of pulsations along the outburst and from spectral analysis \cite[$2 \times 10^8-2 \times 10^{10}$ G,][]{papitto11,miller11}. In particular, our limit is  lower than the magnetic field required ($B_s>10^9$ G) to confine the burning material at the polar cap, as invoked by \cite{cavecchi11} to explain the burst oscillations and their phase locking with the accretion powered pulsations. 

{\bf Acknowlegments}

The author is grateful to Cole Miller and Simon Vaughan for providing insights along the preparation of this paper. The author thank the referee for useful comments. 
\bibliographystyle{apj}

\newpage

\begin{figure}[!t]
\epsscale{.80}
\plotone{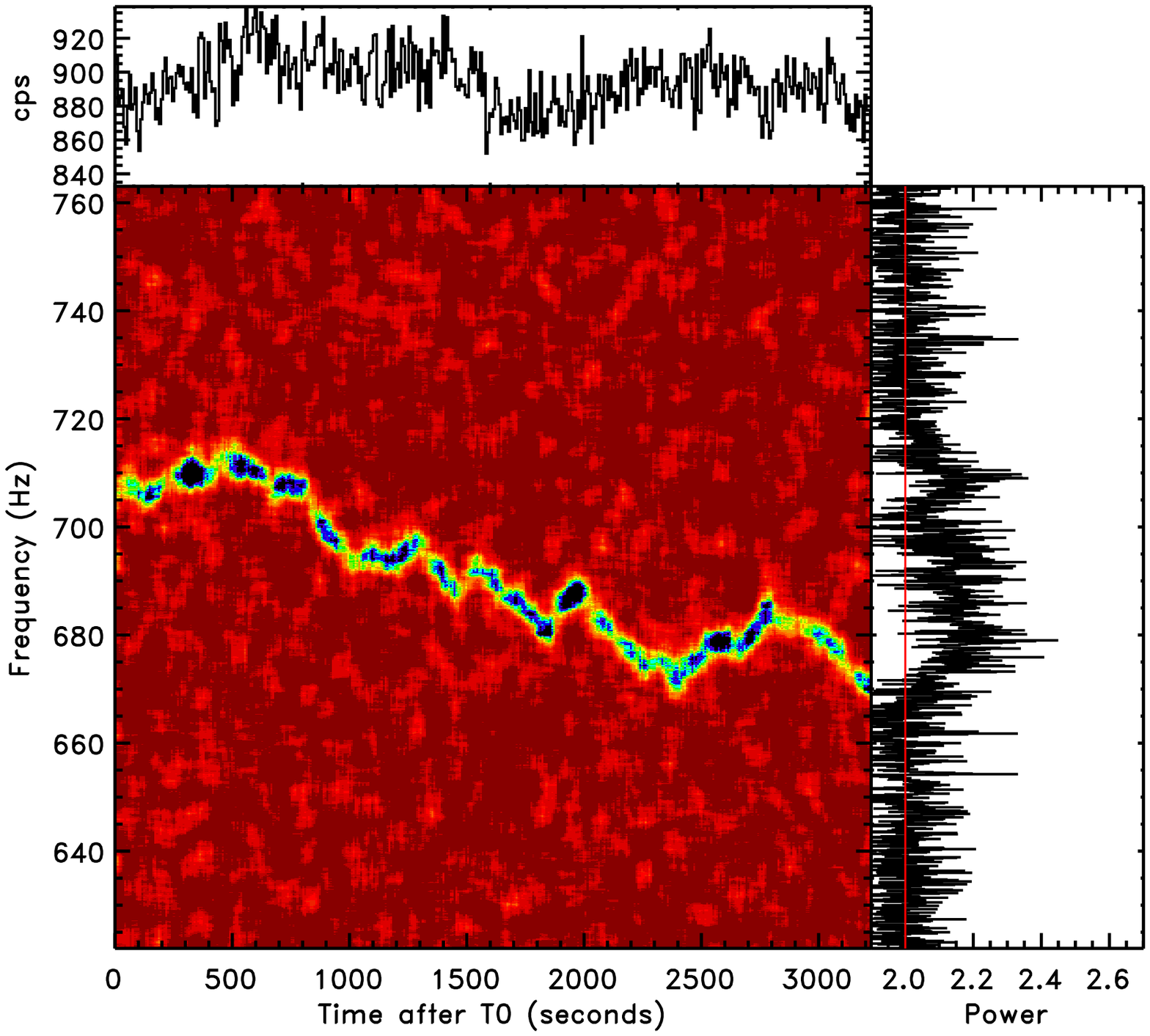}
\caption{Smoothed dynamical PDS for the ObsID 50054-06-11-00 of \exo. The QPO shows big jumps in frequency on hundreds of second timescales. The 3-25 keV light curve at a 8 s resolution is shown at the top, while the projected PDS is plotted on the right hand side panel.}
\label{dynpds}
\end{figure}

\begin{figure}[!t]
\epsscale{.60}
\plotone{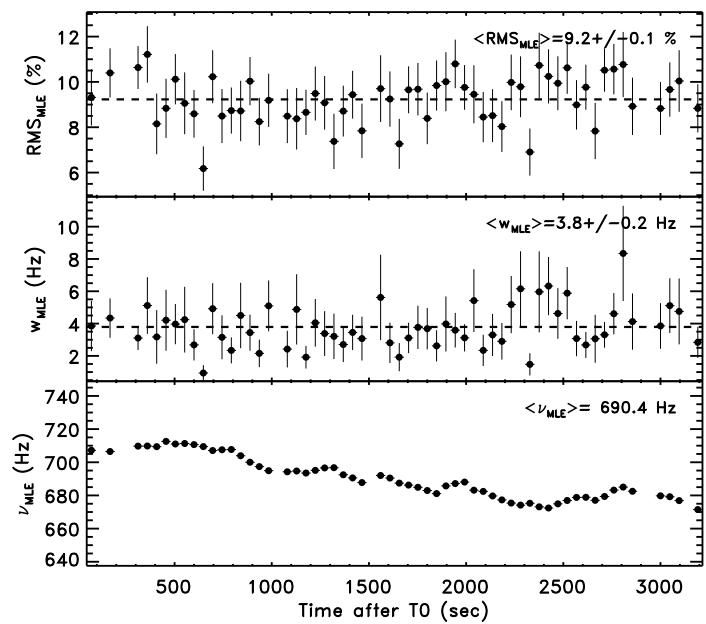}
\caption{Maximum likelihood estimates of the QPO parameters as measured on 48 second timescale (from top to bottom: the fractional RMS amplitude in \%, the QPO width and the QPO frequency). The error bars are given at $1\sigma$ and are computed from the Fisher matrix \citep[][and references therein]{bv11}. Gaps indicate segments for which a significant QPO detection was not obtained. The mean of the MLEs for the RMS amplitude and the width are given inside the panels and shown with a dashed line.}
\label{mle}
\end{figure}
\begin{figure}[!t]
\epsscale{.60}
\plotone{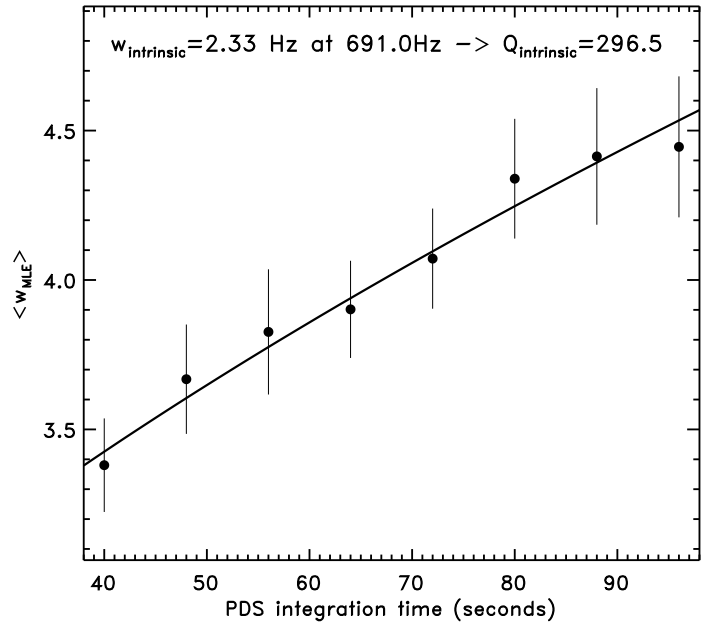}
\caption{Fitting the increase of the measured QPO width with the number of 8 s PDS averaged or equivalently the segment duration with a random walk model (solid line). The intrinsic QPO width and quality factor recovered are 2.3 Hz and 296.5 respectively.}
\label{rw}
\end{figure}
\begin{figure}[!t]
\epsscale{.60}
\plotone{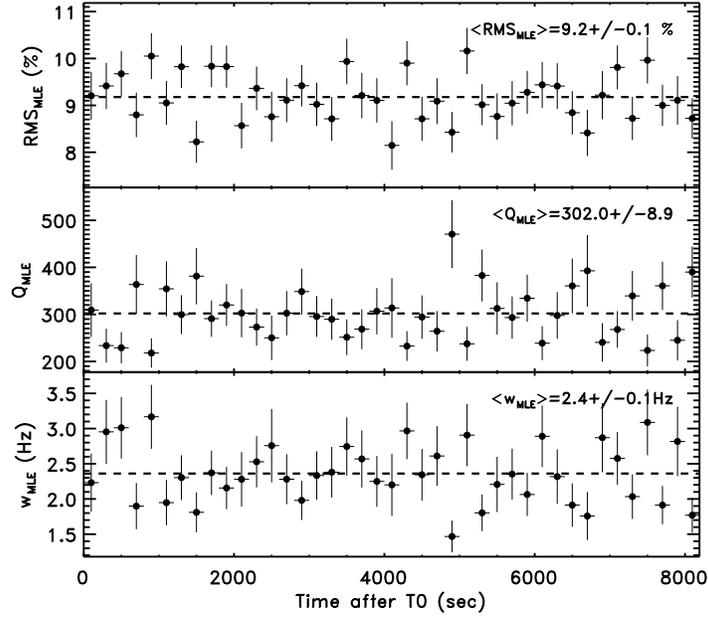}
\caption{Maximum likelihood estimates of the QPO parameters recovered on 200 s duration from a simulation, in which the width, frequency and amplitude of the QPO are constant ($\nu_0=690$ Hz, RMS$_0=9.2$\%, $w_0=2.3$ Hz, equivalent to $Q_0=300$). From top to bottom: the fractional RMS amplitude in \%, the quality factor and the QPO width. The error bars are given at $1\sigma$. This simulation illustrates the typical scattering expected from the $Q$ values measured on this timescale. The mean of the MLEs are given inside the panels and shown with a dashed line.}
\label{mle_sim}
\end{figure}

\begin{figure}[!t]
\epsscale{.60}
\plotone{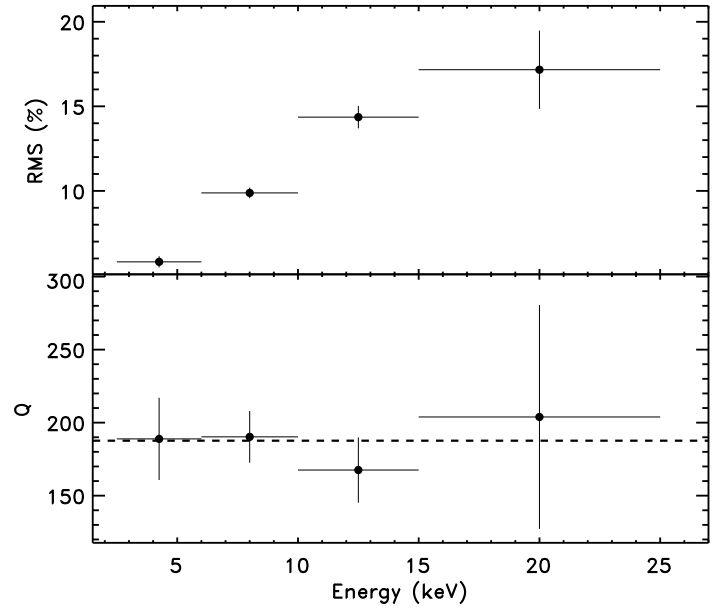}
\caption{Maximum likelihood estimates of the QPO parameters, as derived after shifting-and-adding 400 8 s PDS in four adjacent energy bands. The mean of value of $Q$ is shown with an horizontal dashed line. The statistics of the data does not allow to see variations of $Q$ with energy. }
\label{qene}
\end{figure}

\begin{figure}[!t]
\epsscale{.60}
\plotone{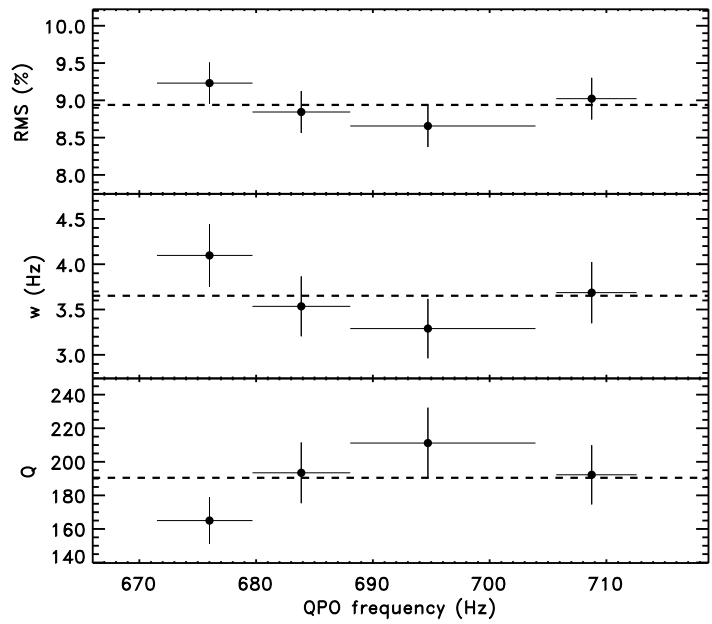}
\caption{Maximum likelihood estimates of the QPO parameters, as derived after shifting-and-adding four sets of 100 8 s PDS, in 4 adjacent frequency intervals. From top to bottom, the RMS amplitude, the width and the quality factor $Q$. The statistics of the data does not allow to see variation of $Q$ with frequency. The mean of the MLEs are shown with a dashed line.}
\label{sa100}
\end{figure}

\begin{table}
\begin{center}
\begin{tabular}{cccccccc}
\multicolumn{7}{c}{ObsID: 95437-01-06-000} \\
\hline
\hline
 Time & T$_{PDS}$ & $\nu$ (Hz) & $w$ (Hz) & $Q$ & RMS (\%) & $R/\delta R$ \\
\hline
2010/10/18 at 07:19:28&1024&803.0$\pm$1.7&11.5$\pm$5.1&70.1$\pm$31.2&3.2$\pm$0.5&3.1\\
2010/10/18 at 07:36:40&1024&828.9$\pm$10.1&40.3$\pm$16.1&20.6$\pm$8.2&4.1$\pm$0.8&2.6\\
2010/10/18 at 07:53:52&1024&808.6$\pm$5.2&42.7$\pm$16.3&18.9$\pm$7.2&4.9$\pm$0.8&3.1\\
\hline
\end{tabular}
\begin{tabular}{cccccccc}
\multicolumn{7}{c}{ObsID: 95437-01-07-00} \\
\hline
\hline
 Time & T$_{PDS}$ & $\nu$ (Hz) & $w$ (Hz) & $Q$ & RMS (\%) & $R/\delta R$ \\
\hline
2010/10/19 at 05:33:54&1024&829.2$\pm$4.3&24.3$\pm$10.6&34.2$\pm$14.9&3.7$\pm$0.7&2.8\\
2010/10/19 at 05:51:06&1024&814.4$\pm$4.8&36.1$\pm$10.6&22.6$\pm$6.7&4.9$\pm$0.7&3.8\\
\hline
\end{tabular}
\begin{tabular}{cccccccc}
\multicolumn{7}{c}{ObsID: 95437-01-09-00} \\
\hline
\hline
 Time & T$_{PDS}$ & $\nu$ (Hz) & $w$ (Hz) & $Q$ & RMS (\%) & $R/\delta R$ \\
\hline
2010/10/21 at 15:20:54&512&851.1$\pm$8.0&46.5$\pm$16.6&18.3$\pm$6.5&6.3$\pm$1.1&2.9\\
2010/10/21 at 15:29:26&512&840.5$\pm$2.9&18.2$\pm$7.3&46.2$\pm$18.6&5.1$\pm$0.8&3.3\\
2010/10/21 at 15:37:58&512&857.4$\pm$1.8&16.7$\pm$6.3&51.3$\pm$19.2&5.4$\pm$0.7&3.7\\
2010/10/21 at 15:46:30&512&858.3$\pm$1.8&10.1$\pm$4.7&84.6$\pm$39.0&3.9$\pm$0.7&2.9\\
2010/10/21 at 15:55:02&512&848.4$\pm$8.2&43.0$\pm$19.3&19.7$\pm$8.8&5.5$\pm$1.1&2.4\\
2010/10/21 at 16:03:42&512&851.8$\pm$2.0&19.3$\pm$4.6&44.2$\pm$10.5&6.4$\pm$0.6&5.3\\
2010/10/21 at 16:12:14&296&863.6$\pm$1.1&5.6$\pm$3.0&155.6$\pm$85.4&3.7$\pm$0.6&2.9\\
\hline
\end{tabular}
\begin{tabular}{cccccccc}
\multicolumn{7}{c}{ObsID: 95437-01-10-01} \\
\hline
\hline
 Time & T$_{PDS}$ & $\nu$ (Hz) & $w$ (Hz) & $Q$ & RMS (\%) & $R/\delta R$ \\
\hline
2010/10/23 at 03:25:60&1024&872.4$\pm$2.0&12.0$\pm$4.1&72.4$\pm$24.9&6.2$\pm$0.8&3.8\\
2010/10/23 at 03:43:04&1024&874.7$\pm$6.6&24.4$\pm$9.4&35.9$\pm$13.8&6.4$\pm$1.1&2.9\\
\hline
\end{tabular}
\begin{tabular}{cccccccc}
\multicolumn{7}{c}{All intervals combined} \\
\hline
\hline
 Time & T$_{PDS}$ & $\nu$ (Hz) & $w$ (Hz) & $Q$ & RMS (\%) & $R/\delta R$ \\
\hline
N/A &10536&839.2$\pm$1.0&25.5$\pm$2.9&32.9$\pm$3.7&4.7$\pm$0.2&11.4\\
\hline
\end{tabular}
\end{center}
\caption{Maximum likelihood estimates of the high frequency QPOs detected from \igr. The data are listed by ObsIDs. The starting time of the Fourier PDS, its cumulative integration time, the QPO frequency,  width,  quality factor, and fractional RMS amplitude of the QPO are given, together with a measure of the significance of the detection, which is given as the ratio of the Lorentzian normalization divided by its $1\sigma$ error (R$\gsim 3$ corresponds to a very significant detection).}
\end{table}

\end{document}